\documentclass{article}
\usepackage{waspaa19,amsmath,graphicx,url,times}
\usepackage{booktabs}
\usepackage{color}
\usepackage{microtype}
\usepackage{xspace}
\usepackage{cite}
\usepackage{dingbat}
\usepackage{balance}

\title{Parametric Resynthesis with neural vocoders}

\twoauthors
 {Soumi Maiti}
    {The Graduate Center, CUNY\\
Computer Science\\
    New York, NY, USA \\
    smaiti@gradcenter.cuny.edu}
 {Michael I Mandel}
    {Brooklyn College \\
     Computer and Information Science\\
    Brooklyn, NY, USA \\
    mim@sci.brooklyn.cuny.edu}

\begin{document}

\ninept
\maketitle

\begin{sloppy}

\begin{abstract}
Noise suppression systems generally produce output speech with compromised quality.
We propose to utilize the high quality speech generation capability of neural vocoders for noise suppression. We use a neural network to predict clean mel-spectrogram features from noisy speech and then compare two neural vocoders, WaveNet and WaveGlow, for synthesizing clean speech from the predicted mel spectrogram.  Both WaveNet and WaveGlow achieve better subjective and objective quality scores than the source separation model Chimera++. Further, WaveNet and WaveGlow also achieve significantly better subjective quality ratings than the oracle Wiener mask. Moreover, we observe that between WaveNet and WaveGlow, WaveNet achieves the best subjective quality scores, although at the cost of much slower waveform generation.

\end{abstract}


\section{Introduction}
\label{sec:intro}

Traditionally, speech enhancement methods modify noisy speech to make it more like the original clean speech~\cite{WangSupervisedSpeechSeparation2018}. Such modification of a noisy signal can introduce additional distortions in the speech signal. Signal distortions generally occur from two problems, over-suppression of the speech and under-suppression of the noise. In contrast, parametric speech synthesis methods can produce high quality speech from only text or textual information. 
Parametric speech synthesis methods predict an acoustic representation of speech from text and then use a vocoder to generate clean speech from the predicted acoustic representation. 

We propose combining speech enhancement and parametric synthesis methods by generating clean acoustic representations from noisy speech and then using a vocoder to synthesize ``clean'' speech from the acoustic representations. We call such a system parametric resynthesis (PR). The first part of the PR system removes noise and predicts the clean acoustic representation. The second part, the vocoder, generates clean speech from this representation. As we are using a vocoder to resynthesize the output speech, the performance of the system is limited by the vocoder synthesis quality.

In our previous work~\cite{maiti2019speech}, we built a PR system with a non-neural vocoder, WORLD~\cite{morise2016world}. Compared to such non-neural vocoders, neural vocoders like WaveNet~\cite{van2016wavenet} synthesize higher quality speech, as shown in the speech synthesis literature~\cite{van2016wavenet, ping2017deep, tamamori2017speaker, wang2017tacotron, shen2017natural, oord2017parallel}. More recent neural vocoders like WaveRNN~\cite{kalchbrenner2018efficient}, Parallel WaveNet~\cite{oord2017parallel}, and WaveGlow~\cite{prenger2018waveglow} have been proposed to improve the synthesis speed of WaveNet while maintaining its high quality. Our goal is to utilize a neural vocoder to resynthesize higher quality speech  from noisy speech  than WORLD allows. We choose WaveNet and WaveGlow for our experiments, as these are the two most different architectures.

In this work we build PR systems with two neural vocoders (PR-neural). Comparing PR-neural to other systems, we show that neural vocoders produce both better speech quality and better noise reduction quality in subjective listening tests than our previous model, PR-World. We show that the PR-neural systems perform better than a recently proposed speech enhancement system, Chimera++ \cite{chimera18}, in all quality and intelligibility scores. And we show that PR-neural can achieve higher subjective intelligibility and quality ratings than the oracle Wiener mask. We also discuss end-to-end training strategies for the PR-neural vocoder system.

\section{Background}
\label{sec:background}
 Speech synthesis can be divided into two broad categories, concatenative and parametric speech synthesis. Traditionally, concatenative speech synthesis has produced the best quality speech. Concatenative systems stitch together small segments of speech recordings to generate new utterances. We previously proposed speech enhancement systems using concatenative synthesis techniques~\cite{maiti2017concatenative, maiti2018large, syed2018concatenative}, named ``concatenative resynthesis.'' Concatenative speech enhancement systems can generate high quality speech with a slight loss in intelligibility, but they are speaker-dependent and generally require a very large dictionary of clean speech. 
 
 With the advent of the WaveNet neural vocoder, parametric speech synthesis with WaveNet surpassed concatenative synthesis in speech quality~\cite{van2016wavenet}. Hence, here we use WaveNet and WaveNet-like neural vocoders for better quality synthesis. A modified WaveNet model, previously has been used as an end-to-end speech enhancement system~\cite{rethage2018wavenet}. This method works in the time domain and models both the speech and the noise present in an observation. Similarly, the SEGAN~\cite{pascual2017segan} and Wave-U-Net~\cite{macartney2018improved} models are end-to-end source separation models that work in the time domain. Both SEGAN and Wave-U-Net down-sample the audio signal progressively in multiple layers and then up-sample them to generate speech. SEGAN which follows a generative adverserial approach has a slightly lower PESQ than Wave-U-Net.   Compared to the WaveNet denoising model of \cite{rethage2018wavenet} and Wave-U-Net, our proposed model is simpler and noise-independent because it does not model the noise at all, only the clean speech.  Moreover, we are able to use the original WaveNet model directly without the modification of \cite{rethage2018wavenet}.

\begin{figure}
  \centering
  \centerline{\includegraphics[width=\linewidth]{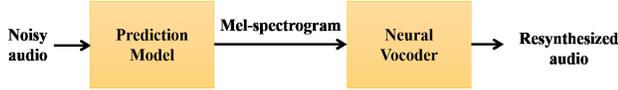}}
  \caption{Parametric Resynthesis model}
  \label{fig:model}
\end{figure}

\section{Model Overview}
\label{sec:model_ovw}

Parametric resynthesis consists of two parts, as shown in Figure~\ref{fig:model}. The first part is a prediction model that predicts the acoustic representation of clean speech from noisy speech. This part of the PR model removes noise from a noisy observation. The second part of the PR model is a vocoder that resynthesizes ``clean'' speech from these predicted acoustic parameters. 
Here we choose to compare two neural vocoders, WaveNet and WaveGlow. 
Both WaveNet and WaveGlow can generate speech conditioned on a log mel-spectrogram, so the log mel-spectrogram is used as the intermediate acoustic parameters.

\subsection{Prediction Model}
\label{ssec:pred_model}
The prediction model uses the noisy mel-spectrogram, $Y(\omega, t)$, as input and the clean mel-spectrogram, $X(\omega,t)$, from parallel clean speech as ground truth. An LSTM~\cite{hochreiter1997long} with multiple layers is used as the core architecture. The model is trained to minimize the mean squared error between the predicted mel-spectrogram, $\hat{X}(\omega,t)$, and the clean mel-spectrogram. 
\begin{equation} \label{eq:pred_loss}
    \mathcal{L} = \sum_{\omega,t} \| X(\omega,t) - \hat{X}(\omega,t) \|^2
\end{equation}
The Adam optimizer is used as the optimization algorithm for training. At test time, given a noisy mel-spectrogram, a clean mel-spectrogram is predicted.

\subsection{Neural Vocoders}
\label{ssec: neu_voc}
Next, conditioned on the predicted mel-spectrogram, a neural vocoder is used to synthesize de-noised speech. We compare two neural vocoders: WaveNet~\cite{van2016wavenet} and WaveGlow~\cite{prenger2018waveglow}. The neural vocoders are trained to generate clean speech from corresponding clean mel-spectrograms.

\subsubsection{WaveNet}
WaveNet~\cite{van2016wavenet} is a speech waveform generation model, built with dilated causal convolutional layers. The model is autoregressive, i.e. generation of one speech sample at time step $t$ ($x_t$) is conditioned on all previous time step samples ($x_1, x_2,...x_{t-1}$). The dilation of the convolutional layers increases by a factor of 2 between subsequent layers and then repeats starting from 1. Gated activations with residual and skip connections are used in WaveNet. It is trained to maximize the likelihood of the clean speech samples. The normalized log mel-spectrogram is used in local conditioning. 

The output of WaveNet is modelled as a mixture of logistic components, as described in~\cite{oord2017parallel, shen2017natural} for high quality synthesis. The output is modelled as a $K$-component logistic mixture.  The model predicts a set of values $\Theta = \{\pi_i, \mu_i, s_i\}_{i=1}^{K}$, where each component of the distribution has its own parameters ${\mu_i, s_i}$ and the components are mixed with probability $\pi_i$. The likelihood of sample $x_t$ is then
\begin{equation} \label{eq:wavenet_loss}
    P(x_t| \Theta, X) = \sum_{i=1}^K \pi_i \left[\sigma \left(\frac{\tilde{x}_{ti} + 0.5}{s_i} \right) - \sigma \left( \frac{\tilde{x}_{ti} - 0.5}{s_i} \right) \right]
\end{equation}
where $\tilde{x}_{ti} = x_t - \mu_i$ and $P(x_t \mid \Theta, X)$ is the probability density function of clean speech conditioned on mel-spectrogram $X$.

We use a publicly available implementation of WaveNet\footnote{\url{https://github.com/r9y9/wavenet_vocoder}} with a setup similar to tacotron2~\cite{shen2017natural}: 24 layers grouped into 4 dilation cycles, 512 residual channels, 512 gate channels, 256 skip channels, and output as mixture-of-logistics with 10 components. As it is an autoregressive model, the synthesis speed is very slow. The PR system with WaveNet as its vocoder is referred to as PR-WaveNet.

\subsubsection{WaveGlow}
WaveGlow~\cite{prenger2018waveglow} is based on the Glow concept~\cite{kingma2018glow} and has faster synthesis than WaveNet. WaveGlow learns an invertible transformation between blocks of eight time domain audio samples and a standard normal distribution conditioned on the log mel spectrogram. It then generates audio by sampling from this Gaussian density. 

The invertible transformation is a composition of a sequence of individual invertible transformations ($f$), normalizing flows. Each flow in WaveGlow consist of a $1 \times 1$ convolutional layer followed by an affine coupling layer. The affine coupling layer is a neural transformation that predicts a scale and bias conditioned on the input speech $x$ and mel-spectrogram $X$. Let $W_k$ be the learned weight matrix for the $k^{\rm th}$ $1 \times 1$ convolutional layer and $s_j(x,X)$ be the predicted scale value at the $j^{\rm th}$ affine coupling layer. 

For inference, WaveGlow samples $z$ from a uniform Gaussian distribution and applies the inverse transformations ($f^{-1}$) conditioned on the mel-spectrogram ($X$) to get back the speech sample $x$. Because parallel sampling from Gaussian distribution is trivial,  all audio samples are generated in parallel. 
The model is trained to minimize the log likelihood of the clean speech samples $x$, 
\begin{multline} \label{eq:waveglow_loss}
    \ln P(x \mid X) = \ln P(z)
    - \sum_{j=0}^J \ln s_j(x,X) - \sum_{k=0}^K \ln |W_k|
\end{multline}
where $J$ is the number of coupling transformations, $K$ is the number of convolutions, $\ln P(z)$ is the log-likelihood of the spherical Gaussian with variance $\nu^2$ and in training $\nu=1$ is used.  Note that WaveGlow refers to this parameter as $\sigma$, but we use $\nu$ to avoid confusion with the logistic function in \eqref{eq:wavenet_loss}. We use the official published waveGlow implementation\footnote{ \url{https://github.com/NVIDIA/waveglow}} with original setup (12 coupling layers, each consisting of 8 layers of dilated convolution with 512 residual and 256 skip connections). We refer to  the PR system with WaveGlow as its vocoder as PR-WaveGlow.

\subsection{Joint Training}
Since the neural vocoders are originally trained on clean mel spectrograms $X(\omega,t)$ and are tested on predicted mel-spectrogram $\hat{X}(\omega,t)$, we can also train both parts of the PR-neural system jointly. The aim of joint training is to compensate for the disparity between the mel spectrograms predicted by the prediction model and consumed by the neural vocoder. Both parts of the PR-neural systems are pretrained then trained jointly to maximize the combined loss of vocoder likelihood and negative mel-spectrogram squared loss.
These models are referred as PR-$\langle$neural vocoder$\rangle$-Joint. We experiment both with and without fine-tuning these models. 


\section{Experiments}
\label{sec:Results}
For our experiments, we use the LJSpeech dataset~\cite{ljspeech17} to which we add environmental noise from CHiME-3~\cite{BarkerthirdCHiMEspeech2015}.  The LJSpeech dataset contains 13100 audio clips from a single speaker with varying length from 1 to 10 seconds at sampling rate of 22~kHz. 
The clean speech is recorded with the microphone in a MacBook Pro in a quiet home environment. CHiME-3 contains four types of environmental noises: street, bus, pedestrian, and cafe. Note that the CHiME-3 noises were recorded at 16~kHz sampling rate. To mix them with LJSpeech, we synthesized white Gaussian noise in the 8-11~kHz band matched in energy to the 7-8~kHz band of the original recordings. The SNR of the generated noisy speech varies from $-9$~dB to $9$~dB SNR with an average of 1~dB. We use 13000 noisy files for training, almost 24 hours of data. The test set consist of 24 files, 6 from each noise type. The SNR of the test set varies from $-7$~dB to $6$~dB. 
The mel-spectrograms are created with window size 46.4~ms, hop size 11.6~ms and with 80 mel bins. The prediction model has 3-bidirectional LSTM layers with 400 units each and was trained with initial learning rate 0.001 for 500 epochs with batch size 64.

Both WaveGlow and WaveNet have published pre-trained models on the LJSpeech data. We use these pre-trained models due to limitations in GPU resources (training the WaveGlow model from scratch takes 2 months on a GPU GeForce GTX 1080 Ti). The published WaveGlow pre-trained model was trained for $580$k iterations (batch size 12) with weight normalization~\cite{weight_norm16}.
The pre-trained WaveNet model was trained for $\sim 1000$k iterations (batch size 2). 
The model also uses L2-regularization with a weight of $10^{-6}$. The average weights of the model parameters are saved as an exponential moving average with a decay of 0.9999 and used for inference, as this is found to provide better quality~\cite{shen2017natural}.
PR-WaveNet-Joint is initialized with the pre-trained prediction model and WaveNet. Then it is trained end-to-end for $355$k iterations with batch size 1. Each training iteration takes $\sim 2.31$~s on a GeForce GTX 1080 GPU.
PR-WaveGlow-Joint is also initialized with the pre-trained prediction and WaveGlow models. It was then trained for $150$k iterations with a batch size of 3.  On a GeForce GTX 1080 Ti GPU, each iteration takes  $> 3$~s.
WaveNet synthesizes audio samples sequentially, the synthesis rate is $\sim 95-98$ samples per second or 0.004$\times$ realtime. Synthesizing 1~s of audio at 22~kHz takes $\sim 232$~s. Because WaveGlow synthesis can be done in parallel, it takes $\sim 1$~s to synthesize $1$~s of audio at a 22~kHz sampling rate.

We compare these two PR-neural models with PR-World, our previously proposed model~\cite{maiti2019speech}, where the WORLD vocoder is used and the intermediate acoustic parameters are the fundamendal frequency, spectral envelope, and band aperiodicity used by WORLD~\cite{morise2016world}. Note that WORLD does not support 22~kHz sampling rates, so this system generates output at 16~kHz. We also compare all PR models with two speech enhancement systems. First is the oracle Wiener mask (OWM), which has access to the original clean speech. The second is a recently proposed source separation system called Chimera++\cite{chimera18}, which uses a combination of the deep clustering loss and mask inference loss to estimate masks. We use our implementation of Chimera++, which we verified to be able to achieve the reported performance on the same dataset as the published model.  It was trained with the same data as the PR systems. 
In addition to the OWM, we measure the best case resynthesis quality by evaluating the neural vocoders conditioned on the true clean mel spectrograms.

Following~\cite{rethage2018wavenet, pascual2017segan, macartney2018improved} we compute composite objective metrics SIG: signal distortion, BAK: background intrusiveness and OVL: overall quality as described in \cite{HuLoizou2006, hu2006evaluation}. All three measures produce numbers between 1 and 5, with higher meaning better quality. We also report PESQ scores as a combined measure of quality and STOI~\cite{taal2010short} as a measure of intelligibility. All test files are downsampled to 16 KHz for measuring objective metrics.

We also conducted a listening test to measure the subjective quality and intelligibility of the systems. For the listening test, we choose 12 of the 24 test files, with three files from each of the four noise types. The listening test follows the Multiple Stimuli with Hidden Reference and Anchor (MUSHRA) paradigm~\cite{MUSHRA}. Subjects were presented with 9 anonymized and randomized versions of each file to facilitate direct comparison: 5 PR systems (PR-WaveNet, PR-WaveNet-Joint, PR-WaveGlow, PR-WaveGlow-Joint, PR-World), 2 comparison speech enhancement systems (oracle Wiener mask and Chimera++), and clean and noisy signals. The PR-World files are sampled at 16~kHz but the other 8 systems used 22~kHz. Subjects were also provided reference clean and noisy versions of each file. Five subjects took part in the listening test. They were told to rate the speech quality, noise-suppression quality, and overall quality of the speech from $0-100$, with $100$ being the best. 

Subjects were also asked to rate the subjective intelligibility of each utterance on the same $0-100$ scale. Specifically, they were asked to rate a model higher if it was easier to understand what was being said. We used an intelligibility rating because in our previous experiments asking subjects for transcripts showed that all systems were near ceiling performance. This could also have been a product of presenting different versions of the same underlying speech to the subjects. Intelligibility ratings, while less concrete, do not suffer from these problems.\footnote{All files are available at \url{http://mr-pc.org/work/waspaa19/}}

\section{Results}
\begin{table}[bt]
    \centering
    \footnotesize
\begin{tabular}{lccc| c |  c }
\toprule
Model & SIG & BAK & OVL & PESQ & STOI \\
\midrule
Clean    & 5.0 & 5.0 & 5.0 & 4.50 & 1.00 \\
WaveGlow & 5.0 & 4.1 & 5.0  & 3.81 & 0.98 \\
WaveNet  & 4.9 & 2.8 & 4.0 & 3.05 & 0.94 \\
\midrule
Oracle Wiener & 4.0  & 2.4 & 3.2 & 2.90 & 0.91\\
\midrule
PR-WaveGlow &  3.9 & 2.5 & 3.1 & 2.58 & 0.87\\
PR-WaveNet   &  3.8 & 2.2 & 3.0 & 2.46 & 0.87\\
Chimera++    & 3.7 & 2.1 & 2.8 & 2.44 & 0.86 \\
PR-WaveGlow-Joint  & 3.6  & 2.5 & 2.9 & 2.28 & 0.84 \\
PR-WaveNet-joint & 3.5 & 2.1 & 2.7 & 2.31 & 0.83\\
PR-World    &  2.8 & 2.1 & 2.3 & 1.53 & 0.79\\
Noisy       &  1.9 &  1.9 & 1.7  & 1.58 & 0.74\\
\bottomrule
\end{tabular}
\caption{Speech enhancement objective metrics:  higher is better. Systems in the top section decode from clean speech as upper bounds. Systems in the middle section use oracle information about the clean speech. Systems in the bottom section are not given any oracle knowledge. All systems sorted by SIG.}
\label{tab:obj}
\end{table}

Table~\ref{tab:obj} shows the objective metric comparison of the systems. In terms of objective quality, comparing neural vocoders synthesizing from clean speech, we observe that WaveGlow scores are higher than WaveNet. WaveNet synthesis has higher SIG quality, but lower BAK and OVL. Comparing the speech enhancement systems, both PR-neural systems outperform Chimera++ in all measures. Compared to the oracle Wiener mask, the PR-neural systems perform slightly worse. After further investigation, we observe that the PR resynthesis files are not perfectly aligned with the clean signal itself, which affects the objective scores significantly. Interestingly, with both, PR-$\langle$neural$\rangle$-Joint performance decreases. When listening to the files, the PR-WaveNet-Joint sometimes contains mumbled unintelligible speech and PR-WaveGlow-Joint introduces more distortions.

In terms of objective intelligibility, we observe the clean WaveNet model has lower STOI than WaveGlow. For the STOI measurement as well, both speech inputs need to be exactly time-aligned, which the WaveNet model does not necessarily provide. The PR-neural systems have higher objective intelligibility than Chimera++.  With PR-WaveGlow, we observe that when trained jointly, STOI actually goes down from 0.87 to 0.84. We observe that tuning WaveGlow's $\sigma$ parameter (our $\nu$) for inference has an effect on quality and intelligibility. When a smaller $\nu$ is used, the synthesis has more speech drop-outs. When a larger $\nu$ is used, these drop-outs decrease, but also the BAK score decreases.
We believe that with a lower $\nu$, when conditioned on a predicted spectrogram, the PR-WaveGlow system only generates segments of speech it is confident in, and mutes the rest.

\begin{figure}[t]
  \centering
  \centerline{\includegraphics[width=0.85\columnwidth, trim={1.6 0.75cm
0.9cm 0.65cm},clip]{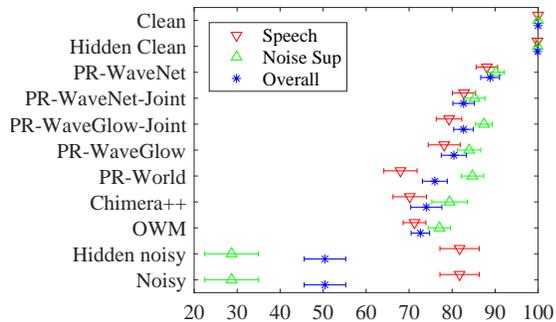}}
  \caption{Subjective quality: higher is better. Error bars show twice the standard error.}
  \label{fig:qual}
\end{figure}

\begin{figure}[t]
  \centering
  \centerline{\includegraphics[width=0.85\columnwidth, trim={1.6 0.7cm
0.85cm 0.5cm},clip]{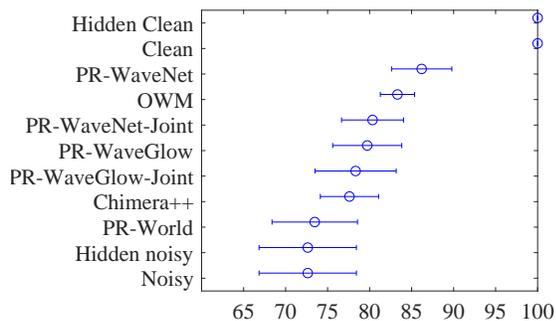}}
  \caption{Subjective Intelligibility: higher is better.}
  \label{fig:intel}
\end{figure}

Figure~\ref{fig:qual} shows the result of the quality listening test. PR-WaveNet performs best in all three quality scores, followed by PR-WaveNet-Joint, PR-WaveGlow-Joint, and PR-WaveGlow. Both PR-neural systems have much higher quality than the oracle Wiener mask. The next best model is PR-WORLD followed by Chimera++. PR-WORLD performs comparably to the oracle Wiener mask, but these ratings are lower than we found in~\cite{maiti2019speech}.  This is likely due to the use of 22~kHz sampling rates in the current experiment but 16~kHz in our previous experiments.

Figure~\ref{fig:intel} shows the subjective intelligibility ratings. We observe that noisy and hidden noisy signals have reasonably high subjective intelligibility, as humans are good at understanding speech in noise. The OWM has slightly higher subjective intelligibility than PR-WaveGlow. PR-WaveNet has slightly but not significantly higher intelligibility, and the clean files have the best intelligibility. The PR-$\langle$neural$\rangle$-Joint models have lower intelligibility, caused by the speech drop-outs or mumbled speech as mentioned above.

\begin{table}[bt]
    \centering
    \footnotesize
\begin{tabular}{l cc ccc| c |  c }
\toprule
& \multicolumn{2}{c}{Fine-tuned} \\
Model & Pred. & Voc. & SIG & BAK & OVL & PESQ & STOI \\
\midrule
WaveNet  & & & 3.8 & 2.2 & 3.0 & 2.46 & 0.87\\
WaveNet & \checkmark & & 3.9 & 2.2 & 3.1 & 2.49 & 0.88 \\
WaveNet & & \checkmark      & 3.1 & 1.9 & 2.3  & 2.02 & 0.78 \\
WaveNet & \checkmark & \checkmark        & 3.5 & 2.1 & 2.7 & 2.29 & 0.83\\
\midrule
WaveGlow & & & 3.9 & 2.5 & 3.1 & 2.58 & 0.87\\
WaveGlow & \checkmark & & 4.0 & 2.5 &  3.2 & 2.70 & 0.90 \\
WaveGlow & & \checkmark & 3.6 & 2.5  & 2.9  & 2.24 & 0.82 \\
WaveGlow & \checkmark & \checkmark & 3.6  & 2.4 & 2.9 & 2.28 & 0.84 \\
\bottomrule
\end{tabular}
\caption{Objective metrics for different joint fine-tuning schemes for PR-neural systems components. }
\label{tab:joint}
\end{table}

\section{Discussion of Joint Training}
Table~\ref{tab:joint} shows the results of further investigation of the drop in performance caused by jointly training the PR-neural systems. The PR-$\langle$neural$\rangle$-Joint models are trained using the vocoder losses. After joint training, both WaveNet and WaveGlow seemed to change the prediction model to make the intermediate clean mel-spectrogram louder. As training continued, this predicted mel-spectrogram did not approach the clean spectrogram, but instead became a very loud version of it, which did not improve performance. When the prediction model was fixed and only the vocoders were fine-tuned jointly, we observed a large drop in performance. In WaveNet this introduced more unintelligible speech, making it smoother but garbled. In WaveGlow this increased speech dropouts (as can be seen in the reduced STOI scores).  
Finally with the neural vocoder fixed, we trained the prediction model to minimize a combination of mel spectrogram MSE and vocoder loss. This provided slight improvements in performance: both PR-WaveNet and PR-WaveGlow improved intelligibility scores as well as SIG and OVL. 

\section{Conclusion}
This paper proposes the use of neural vocoders in parametric resynthesis for high quality speech enhancement. We show that using two neural vocoders, WaveGlow and WaveNet, produces better quality enhanced speech than using a traditional vocoder like WORLD. We also show that PR-neural models outperform the recently proposed Chimera++ mask-based speech enhancement system in all intelligibility and quality scores. Finally we show that PR-WaveNet achieves significantly better subjective quality scores than the oracle Wiener mask. In future, we will explore the speaker-dependence of these models.

\section{Acknowledgements}

This material is based upon work supported by the National Science Foundation (NSF) grant IIS-1618061. Any opinions, findings, and conclusions or recommendations expressed in this material are those of the author(s) and do not necessarily reflect the views of the NSF.

\balance
\bibliographystyle{IEEEtran}
\bibliography{ref}

\begin{thebibliography}{10}
\providecommand{\url}[1]{#1}
\def\UrlFont{\rmfamily}
\providecommand{\newblock}{\relax}
\providecommand{\bibinfo}[2]{#2}
\providecommand\BIBentrySTDinterwordspacing{\spaceskip=0pt\relax}
\providecommand\BIBentryALTinterwordstretchfactor{4}
\providecommand\BIBentryALTinterwordspacing{\spaceskip=\fontdimen2\font plus
\BIBentryALTinterwordstretchfactor\fontdimen3\font minus
  \fontdimen4\font\relax}
\providecommand\BIBforeignlanguage[2]{{%
\expandafter\ifx\csname l@#1\endcsname\relax
\typeout{** WARNING: IEEEtran.bst: No hyphenation pattern has been}%
\typeout{** loaded for the language `#1'. Using the pattern for}%
\typeout{** the default language instead.}%
\else
\language=\csname l@#1\endcsname
\fi
#2}}

\bibitem{WangSupervisedSpeechSeparation2018}
D.~Wang and J.~Chen, ``Supervised speech separation based on deep learning: An
  overview,'' \emph{IEEE/ACM Transactions on Audio, Speech, and Language
  Processing}, vol.~26, no.~10, pp. 1702--1726, Oct. 2018.

\bibitem{maiti2019speech}
S.~Maiti and M.~I. Mandel, ``Speech denoising by parametric resynthesis,''
  \emph{arXiv preprint arXiv:1904.01537}, 2019.

\bibitem{morise2016world}
M.~Morise, F.~Yokomori, and K.~Ozawa, ``{WORLD}: a vocoder-based high-quality
  speech synthesis system for real-time applications,'' \emph{IEICE
  Transactions on Information and Systems}, vol.~99, no.~7, pp. 1877--1884,
  Jul. 2016.

\bibitem{van2016wavenet}
A.~van~den Oord, S.~Dieleman, H.~Zen, K.~Simonyan, O.~Vinyals, A.~Graves,
  N.~Kalchbrenner, A.~W. Senior, and K.~Kavukcuoglu, ``{WaveNet}: A generative
  model for raw audio.'' in \emph{Proc.~ISCA SSW}, Sept. 2016, p. 125.

\bibitem{ping2017deep}
W.~Ping, K.~Peng, A.~Gibiansky, S.~O. Arik, A.~Kannan, S.~Narang, J.~Raiman,
  and J.~Miller, ``Deep voice 3: 2000-speaker neural text-to-speech,''
  \emph{arXiv preprint arXiv:1710.07654}, 2017.

\bibitem{tamamori2017speaker}
A.~Tamamori, T.~Hayashi, K.~Kobayashi, K.~Takeda, and T.~Toda,
  ``Speaker-dependent {WaveNet} vocoder,'' in \emph{Proc. Interspeech}, vol.
  2017, 2017, pp. 1118--1122.

\bibitem{wang2017tacotron}
Y.~Wang, R.~Skerry-Ryan, D.~Stanton, Y.~Wu, R.~J. Weiss, N.~Jaitly, Z.~Yang,
  Y.~Xiao, Z.~Chen, S.~Bengio, \emph{et~al.}, ``Tacotron: A fully end-to-end
  text-to-speech synthesis model,'' \emph{arXiv preprint arXiv:1703.10135},
  2017.

\bibitem{shen2017natural}
J.~Shen, R.~Pang, R.~J. Weiss, M.~Schuster, N.~Jaitly, Z.~Yang, Z.~Chen,
  Y.~Zhang, Y.~Wang, R.~Skerry-Ryan, \emph{et~al.}, ``Natural tts synthesis by
  conditioning wavenet on mel spectrogram predictions,'' \emph{arXiv preprint
  arXiv:1712.05884}, 2017.

\bibitem{oord2017parallel}
A.~v.~d. Oord, Y.~Li, I.~Babuschkin, K.~Simonyan, O.~Vinyals, K.~Kavukcuoglu,
  G.~v.~d. Driessche, E.~Lockhart, L.~C. Cobo, F.~Stimberg, \emph{et~al.},
  ``Parallel wavenet: Fast high-fidelity speech synthesis,'' \emph{arXiv
  preprint arXiv:1711.10433}, 2017.

\bibitem{kalchbrenner2018efficient}
N.~Kalchbrenner, E.~Elsen, K.~Simonyan, S.~Noury, N.~Casagrande, E.~Lockhart,
  F.~Stimberg, A.~v.~d. Oord, S.~Dieleman, and K.~Kavukcuoglu, ``Efficient
  neural audio synthesis,'' \emph{arXiv preprint arXiv:1802.08435}, 2018.

\bibitem{prenger2018waveglow}
R.~Prenger, R.~Valle, and B.~Catanzaro, ``Waveglow: A flow-based generative
  network for speech synthesis,'' \emph{arXiv preprint arXiv:1811.00002}, 2018.

\bibitem{chimera18}
Z.~{Wang}, J.~L. {Roux}, and J.~R. {Hershey}, ``Alternative objective functions
  for deep clustering,'' in \emph{Proc. ICASSP}, Apr. 2018, pp. 686--690.

\bibitem{maiti2017concatenative}
S.~Maiti and M.~I. Mandel, ``Concatenative resynthesis using twin networks,''
  \emph{Proc. Interspeech}, pp. 3647--3651, 2017.

\bibitem{maiti2018large}
S.~Maiti, J.~Ching, and M.~Mandel, ``Large vocabulary concatenative
  resynthesis,'' in \emph{Proc. Interspeech}, 2018.

\bibitem{syed2018concatenative}
A.~R. Syed, T.~V. Anh, and M.~I. Mandel, ``Concatenative resynthesis with
  improved training signals for speech enhancement,'' in \emph{Proc.
  Interspeech}, 2018.

\bibitem{rethage2018wavenet}
D.~Rethage, J.~Pons, and X.~Serra, ``A wavenet for speech denoising,'' in
  \emph{Proc.~ICASSP}, 2018, pp. 5069--5073.

\bibitem{pascual2017segan}
S.~Pascual, A.~Bonafonte, and J.~Serr{\`a}, ``Segan: Speech enhancement
  generative adversarial network,'' \emph{arXiv preprint arXiv:1703.09452},
  2017.

\bibitem{macartney2018improved}
C.~Macartney and T.~Weyde, ``Improved speech enhancement with the wave-u-net,''
  \emph{arXiv preprint arXiv:1811.11307}, 2018.

\bibitem{hochreiter1997long}
S.~Hochreiter and J.~Schmidhuber, ``Long short-term memory,'' \emph{Neural
  Computation}, vol.~9, no.~8, pp. 1735--1780, 1997.

\bibitem{kingma2018glow}
D.~P. Kingma and P.~Dhariwal, ``Glow: Generative flow with invertible 1x1
  convolutions,'' \emph{arXiv preprint arXiv:1807.03039}, 2018.

\bibitem{ljspeech17}
K.~Ito, ``The {LJ} speech dataset,''
  \url{https://keithito.com/LJ-Speech-Dataset/}, 2017.

\bibitem{BarkerthirdCHiMEspeech2015}
J.~Barker, R.~Marxer, E.~Vincent, and S.~Watanabe, ``The third ‘{CHiME}’
  speech separation and recognition challenge: Dataset, task and baselines,''
  in \emph{Proc.~ASRU}, 2015, pp. 504--511.

\bibitem{weight_norm16}
T.~Salimans and D.~P. Kingma, ``Weight normalization: A simple
  reparameterization to accelerate training of deep neural networks,'' in
  \emph{Proc. NIPS}, 2016, pp. 901--909.

\bibitem{HuLoizou2006}
Y.~{Hu} and P.~C. {Loizou}, ``Subjective comparison of speech enhancement
  algorithms,'' in \emph{Proc. ICASSP}, vol.~1, May 2006, pp. I--I.

\bibitem{hu2006evaluation}
Y.~Hu and P.~C. Loizou, ``Evaluation of objective measures for speech
  enhancement,'' in \emph{Proc. Interspeech}, 2006.

\bibitem{taal2010short}
C.~H. Taal, R.~C. Hendriks, R.~Heusdens, and J.~Jensen, ``A short-time
  objective intelligibility measure for time-frequency weighted noisy speech,''
  in \emph{Proc.~ICASSP}, 2010, pp. 4214--4217.

\bibitem{MUSHRA}
``Method for the subjective assessment of intermediate quality level of audio
  systems,'' International Telecommunication Union Radiocommunication
  Standardization Sector ({ITU-R}), Tech. Rep. BS.1534-3, 2015.

\end{thebibliography}

\end{sloppy}
\end{document}